# A Boltzmann Machine Implementation for the D-Wave

John E. Dorband Ph.D.
Dept. of Computer Science and Electrical Engineering
University of Maryland, Baltimore County
Baltimore, Maryland 21250, USA
dorband@umbc.edu

*Abstract*—The D-Wave is an adiabatic quantum computer. It is an understatement to say that it is not a traditional computer. It can be viewed as a computational accelerator or more precisely a computational oracle, where one asks it a relevant question and it returns a useful answer. The question is how do you ask a relevant question and how do you use the answer it returns. This paper addresses these issues in a way that is pertinent to machine learning. A Boltzmann machine is implemented with the D-Wave since the D-Wave is merely a hardware instantiation of a partially connected Boltzmann machine. This paper presents a prototype implementation of a 3-layered neural network where the D-Wave is used as the middle (hidden) layer of the neural network. This paper also explains how the D-Wave can be utilized in a multi-layer neural network (more than 3 layers) and one in which each layer may be multiple times the size of the D-Wave being used.

## I. Introduction

The D-Wave is an adiabatic quantum computer [1][2]. Its function is to determine a set of ones and zeros (qubits) that minimize the objective function, equation (1).

$$E = \sum_i a_i q_i + \sum_i \sum_j b_{ij} q_i q_j \qquad (1)$$

where E is the objective function (energy), $q_i$ are the qubits and the $a_i$ and the $b_{ij}$ are the coefficients that the programmer provides to the D-wave that describes the problem the programmer wishes to be solved. The D-Wave solves a restricted form of the above expression, one that can be mapped to a Chimera graph. D-Wave's SYSTEM6 of which most of this work was performed on consists of 64 (8x8) cells of 8 qubits each arranged as a chimera graph (see figure 1).

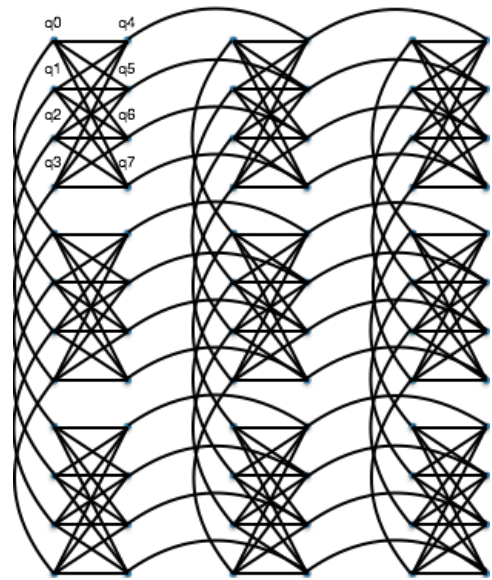

Figure 1. Nine cells of a chimera graph. Each cell contains 8 qubits.

Note that the 8 qubits of a cell form a bipartite graph of the 4 lower numbered qubits and the 4 higher numbered qubits. Various problem solutions may be formulated with the proper choice of coefficients ($a_i$ and $b_{ij}$) that are mapped on to the appropriate chimera graph. Such applications are logic circuit simulation (e.g. an integer multiplier), solution existence problems like NP-complete problems (eq. SAT, traveling salesman), or optimization problems. This paper presents another class of algorithms performed on the D-Wave, those formulated to utilize neural networks. First, an attempt is made to characterize the D-Wave in terms of stochastic qualities.

## II. Qubit Computational Characteristics

An experiment was performed to determine the stochastic characteristics of the qubits. All the coupling coefficients were set to zero. This decouples the qubits from each other. All the qubits' coefficients were set to the same value. The D-Wave was requested to run with these coefficients for 10,000 tries.

The number of times each qubit had a value of one (true) was count over the 10,000 tries and divided by 10,000 to determine a probability of a specific qubit of having a value of one given the coefficient values for that set of tries. The common qubit coefficient value was varied from -1 to 1 in increments of 1/64. Figure 2 shows the results of this test.

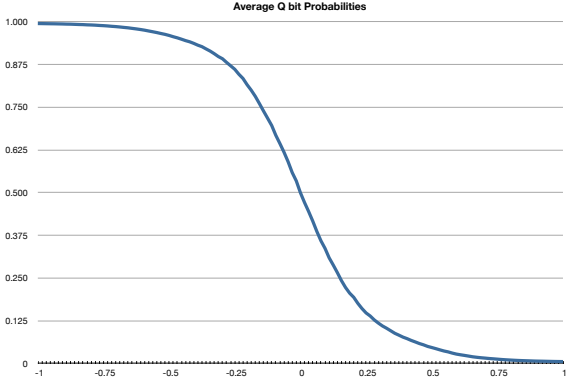

Figure 2. The probability of a decoupled qubit of having a value of 1 given a specific coefficient value.

A qubit coefficient value of zero gives extremely close to a 50/50 chance of the qubit being a zero or a one. These values are an average over all qubits. Note that the measurements of figure 2 form a sigmoid probability distribution. In fact this distribution is extremely close to $P(b) = 1/(1 + e^{7b})$.

If one looks at the qubits they varies from qubit to qubit. Figure 3 shows the standard deviation across the qubits for each common qubit coefficient. The variation is greatest near zero because there is a greater freedom to vary in the middle of the common coefficient range from -1 to 1.

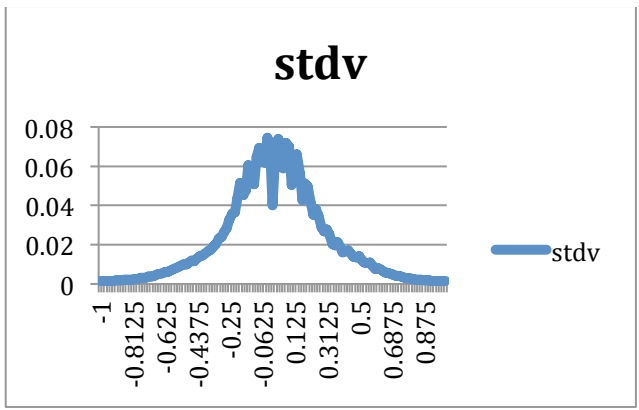

Figure 3. The standard deviation of the probability of a qubit being one over all qubits for each common qubit coefficent value.

### III. QUBIT COUPLING CHARACTERISTICS

The metric for coupling is defined next. Let $Q_T^A$ be the probability that qubit $A$ has a value $T$ where $T$ is either 0 or 1. And $Q_{TS}^{AB}$ represents the probability that qubit $A$ has a value $T$ while $B$ has a value of $S$. We know that if $Q_T^A$ is independent of $Q_S^B$ then $Q_{TS}^{AB} = Q_T^A Q_S^B$. That is, $Q_{TS}^{AB}$ is factorable. Let $Q_1^A = a$, $Q_0^A = b$, $Q_1^B = c$, and $Q_0^B = d$. Then $Q_{11}^{AB} = U = ac$, $Q_{10}^{AB} = V = ad$, $Q_{01}^{AB} = W = bc$, and $Q_{00}^{AB} = X = bd$. Therefore, $\frac{U}{W} = \frac{a}{b}$ and $\frac{V}{X} = \frac{a}{b}$ implies $UX = VW = abcd$. If qubits $A$ and $B$ are independent (uncoupled), then $UX = VW$. That is $Q_{11}^{AB} Q_{00}^{AB}$ must equal $Q_{01}^{AB} Q_{10}^{AB}$. If however $Q_{11}^{AB} Q_{00}^{AB} \neq Q_{01}^{AB} Q_{10}^{AB}$, then qubits $A$ and $B$ must be coupled. The following will be used as a metric of coupling:

$$coupling = log\left(\frac{Q_{11}^{AB} Q_{00}^{AB}}{Q_{01}^{AB} Q_{10}^{AB}}\right) \quad (2)$$

If the coupling is zero the qubits are uncoupled and if the coupling is not equal to zero, they must be coupled. Figure 4 shows the experimentally measured probability of two paired qubits of having a value of either 1 and 1, 1 and 0, 0 and 1, or 0 and 0, averaged of the 222 tested qubit pairs. Figure 5 is the coupling factor calculated from the measured data.

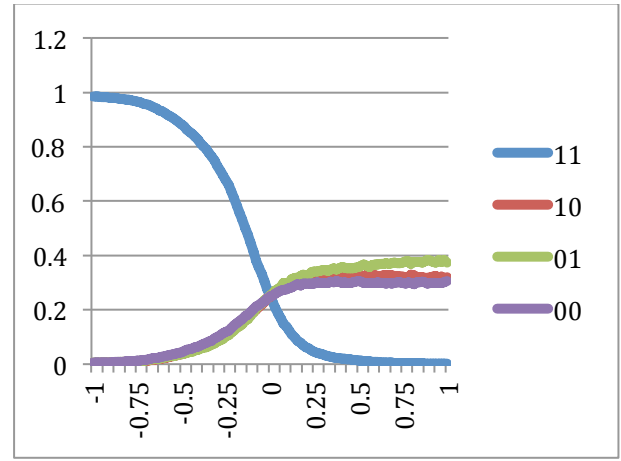

Figure 4. The probability of two qubits having a value of 1&1, 1&0, 0&1, or 0&0.

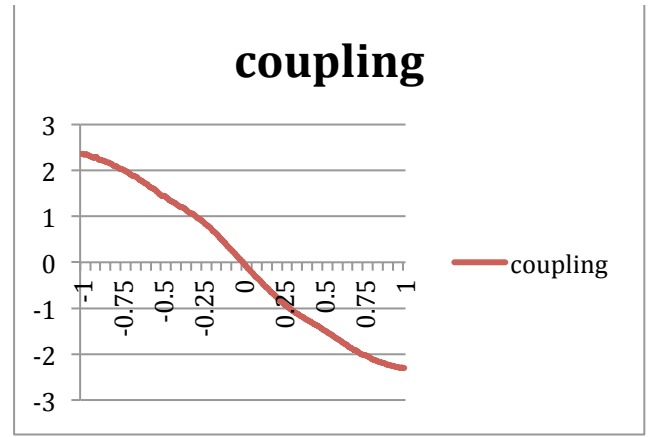

Figure 5. The coupling of qubit averaged over 222 qubit pairs, based on data from figure 4.

The results of an experiment where 222 pairs of qubits are used to test the characteristics of the coupling of qubits. The coefficient of each qubit is set to zero and the value of the coupling coefficient is varied from -1 to 1.

These stochastic characteristics of the D-Wave qubits and the way the qubits are coupled are precisely what is necessary for the D-Wave to fit into some models of neural network implementations, in particular Hopfield networks and Boltzmann machines.

## IV. Hopfield Networks and Boltzmann Machines.

In 1982, the Hopfield network was invented by John Hopfield[3]. This network is a recurrent artificial neural network. Its purpose was to implement content addressable memory. The nodes of the network were binary thresholding units. A binary thresholding unit has two states, either one or zero depending on a thresholding method or in transition between those two states. The connections between the binary units are bidirectional. That is the coupling values of the binary units equally effect the stochastic behavior of both coupled binary units. For a Hopfield network the thresholding method can be either stable, oscillatory, or chaotic. The Hopfield network is based on the concept of minimization of energy where energy is defined in equation (3).

$$E = \sum_i b_i s_i + \sum_i \sum_j w_{ij} s_i s_j \quad (3)$$

where the $w$'s are the connection weights, the $b$'s are the binary unit biases, and the $s$'s are the values of the binary units. Note that the same expression is used to describe the objective function, equation (1), that is minimized by the D-Wave architecture.

In 1985, Geoffrey Hinton and Terry Sejnowski[4][5] invented the Boltzmann machine. It is a Hopfield network with stochastic binary thresholding units. Again noting that the D-Wave qubits are effectively binary stochastic units. A Boltzmann machine as with a Hopfield network consists of a completely connected graph. It is extremely computational intensive to compute the minimum energy of such networks, if not intractable for other than trivial graphs. So an alternative network was developed by Geoffrey Hinton, the restricted Boltzmann machine. With this network there are no connections between binary units within a layer, yet the units between two connected layers form a completely connected bipartite graph. This dramatically reduces the computation of the minimum energy of a multilayer neural network.

In 2005, Carreira-Perpinan and Hinton [6] developed an efficient method for updating connection weights on a Boltzmann Machine called contrastive divergence. A variation on this method was used in the implementation of the neural network using the D-Wave discussed in this paper.

More recently work has been done by Denil and de Freitas[7] and Dumoulin et al. [8] to simulate in software a Boltzmann machine running on a D-Wave machine.

## V. Implementing a Neural Network on the D-Wave.

An artificial neural network consists of artificial neurons. Artificial neurons are made of connections and thresholding nodes. The connections transfer values to and from nodes. The values are then accumulated on the nodes from incoming connections and thresholded. The thresholded value is sent out over outgoing connections. Connections can either be uni-directional or bi-directional (symmetric). The prototype implemented here will be based on figure 6, but only contained 3 layers. Figure 6 is a basic neural network with 5 layers, 2 visible layers and 3 hidden layers, containing various numbers of binary units in each layer and couplers as connections between units.

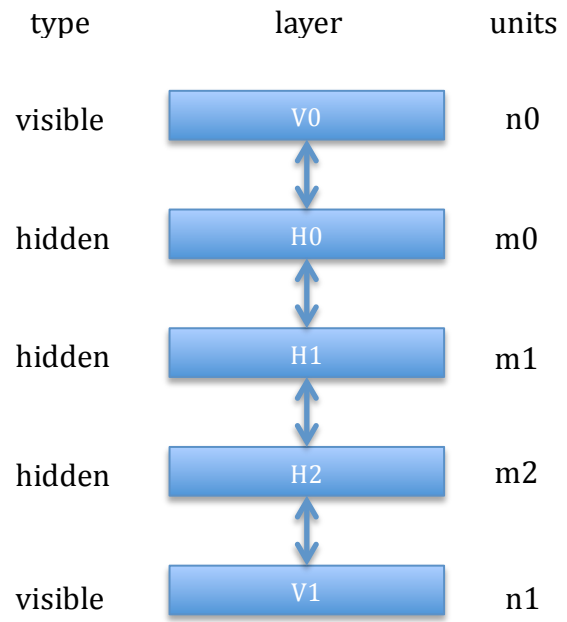

Figure 6. A example neural network

It will be shown in this paper how a D-Wave may be utilized to implement a multi-layered neural network where each hidden layer of the neural network is a partially connected Boltzmann machine. A partially connected Boltzmann machine has some connections between units within a layer, in this case that of D-Wave, a Chimera graph. The connections between layers form completely connected bipartite graphs. There are no connections between units in the visible layers. Layers are coupled to each other by calculating a bias value for each binary unit in a layer. The bias value of a binary unit consists of the average bias values from all binary units coupled to that binary unit that are not in the same layer as that binary unit. For example a binary unit in layer H1 is coupled to m0 binary units in H0 and m2 binary units in H2. For each binary unit in H0 and H2 a bias contribution $b_i$ is compute for each binary unit, $i$, in H1.

$b_i = w_{ij} e_j$, where $w_{ij}$ is the coupling coefficient for the coupler between binary units $i$ and $j$, and $e_j$ is the expected value for the binary unit in layers H0 and H2. For any given binary unit in H1 an average bias contribution of all connected binary unit in H2 is computed. Then the bias contributions from H2 and H0 are averaged together. This become the bias of that binary unit in H1. In this way the bias contribution from H0 is given the same weighting as the bias from H2 no matter the difference of the size of m0 and m1. Note that the coupling coefficients between layers are not used by or sent to the D-Wave and coupling coefficients used by the D-Wave are not used in computing the D-Wave binary unit (qubit) contributions to the binary unit's bias.

If a D-Wave larger than physically exists is needed, a virtual D-Wave can be configured using the above described technique. Virtual couplers are defined along the edges of the physical D-Wave to couple the physical qubits of the D-Wave. Thus for a virtual D-Wave neural network layer twice the size of the physical D-Wave, the physical D-Wave would need to be run twice and the virtual coupler coefficients between the two parts of the virtual D-Wave would be updated at the same time the coupler between different D-Wave layers of the neural network. This could be extended to any size virtual D-Wave neural network layer.

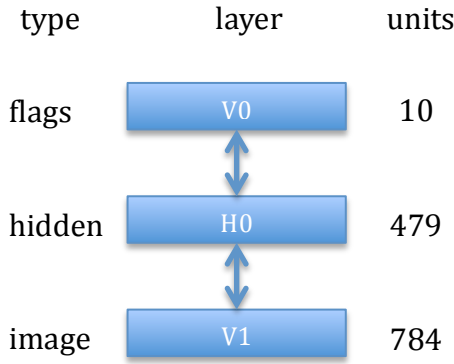

Figure 7. The D-Wave implemented neural network for recognizing handwritten digits.

As a prototype, a 3 layer neural network was implemented to recognize handwritten digits. The MNIST handwritten digit dataset was used to train and test the neural network. This supervised neural network (figure 7) consisted of 1 hidden layer (479 D-Wave qubits) and 2 visible layers, the digit image (28x28) layer and the digit indicator flag layer (10). The connections between layers form completely connected bipartite graphs. There are no connections between units in the visible layers. The network is treated as though it were executing in parallel. All the interactions of the layers are set up for execution, then each layer is processed independent of the others. This style of performance is continued through out the teaching of the network as well as the testing. This style of processing allows for the greatest degree of independence between steps of the processing. During teaching the network connection weights are computed using a three phase update process, setup, execution and update. Phase 1 of the processing, process set up, primarily involves computing binary unit biases, phase 2 computes expected values of the binary units and the third phase is to update the connection weights. Phase 1 and 2 maybe iterated several times (in this case twice) before phase 3 is performed. In the case of D-Wave the biases are use to couple the D-Wave to the other layers of the network. This is done for each D-Wave stochastic binary unit (qubit) by setting the coefficient of the qubit ($b_i$) to the average product of the expected value ($e_j$) of the binary units connected to qubit and the weight of that connection ($w_{ij}$).

$$b_i = log(w_{ij}e_j) \quad (4)$$

The reason to use the average rather than the sum is that the values to which a qubit coefficient on the D-Wave will be set is between -1 and 1.

During phase 2 for the D-Wave, the D-Wave is sent both the qubit coefficients and the appropriate network weights (qubit coupling coefficients). One thousand sets of results are requested from the D-Wave. The D-wave returns the results of the request and the results are processed to determine the expected values for the D-Wave qubits. For binary units not evaluated on the D-Wave, those in the visible layers, phase 1 consists of once again taking an average of the inputs from connected binary units resulting in a bias. And phase 2 consists of just applying the sigmoid function that was obtained from characterizing the D-Wave qubit behavior to that bias resulting in an expected value for that binary unit.

Initially, the image binary units are set to normalized image values between 0 and 1 corresponding to the brightness value of each pixel of the image and the flag value is set to 1 for the flag that corresponds to image type (0-9) and zero for all other flags. The hidden units' expected values are initialized to values that correspond to value the D-Wave qubit would have if all qubit coefficients had been set to 0. A good approximation to this is to set all qubits' expected values to 0.5, but since the actual values have been measures, the measured values are actually used. In the case of a 3 layered network this is unimportant since the qubits' expected values aren't actually used, but in the case of networks with more layers they would need to be used.

Phase 3 is the update phase. Phase 1 and 2 are run 2 or more times in a series before phase 3 is run. The expected values, $f_i$, are saved after the first run of phases 1 and 2. The expected values, $e_i$, are saved after the last execution of phase 1 and 2 of a series. These values are use in the following expression to update the weights (coupling coefficients, $w_{ij}$)

$$\Delta w_i = k(e_i e_j - f_i f_j) \quad (5)$$

where k controls the learning rate.

Here is a quick summary of a training session of the neural network. A training session consists of multiple passes over the training set. The training set consists of multiple 28x28 pixel images of hand written digits, 0 to 9. Each pass over the training set consists of one processing cycle per image. A processing cycle consists of initializing the neural network with an image and executing phase 1 and 2 multiple times followed by a phase 3.

A test cycle consists of initializing the image layer of the neural network with a digit image and initializing all the flags to zero. The phase 1 and phase 2 are run once. This will compute expected values for the flags. The flag with the highest expected value is considered the neural networks first choice, the flag with the second highest expected value is considered the second choice, and the third highest, the third choice.

## VI. RESULTS OF TEACHING MNIST ON THE D-WAVE.

The neural network was trained and tested on three training/test sets, one with 50 training images and 50 test images, one with 100 training and 100 test, and one with 200 training and 200 test. The digit images were selected from the 10,000 image MNIST image data set. The training sets also contained gray (random noise) images for which all flags were set to zero. The three training sets contained 10, 20, and 40 gray images respectively. Tables I through III contain the results of the neural network trained with the three training sets. Each trained neural network was tested with the set it was trained with as well as a different test set. The tables contain the probabilities in which the network recognized the digit correctly by its first, second, or third choice.

TABLE I. RECOGNITION PROBABILITY FOR 50 TRAINING IMAGES AND 50 TEST IMAGES.

| Passes | Training Set | | | Test Set | | |
|---|---|---|---|---|---|---|
| | 1st Choice | by 2nd | by 3rd | 1st Choice | by 2nd | by 3rd |
| 10 | .90 | 1.00 | 1.00 | .62 | .80 | .86 |
| 20 | .96 | .98 | 1.00 | .66 | .80 | .88 |
| 30 | .98 | .98 | 1.00 | .70 | .80 | .86 |

TABLE II. RECOGNITION PROBABILITY FOR 100 TRAINING IMAGES AND 100 TEST IMAGES.

| Passes | Training Set | | | Test Set | | |
|---|---|---|---|---|---|---|
| | 1st Choice | by 2nd | by 3rd | 1st Choice | by 2nd | by 3rd |
| 10 | .84 | .95 | .99 | .59 | .71 | .81 |
| 20 | .91 | .98 | .98 | .56 | .68 | .75 |
| 30 | .93 | .97 | 1.00 | .56 | .72 | .81 |

TABLE III. RECOGNITION PROBABILITY FOR 200 TRAINING IMAGES AND 200 TEST IMAGES.

| Passes | Training Set | | | Test Set | | |
|---|---|---|---|---|---|---|
| | 1st Choice | by 2nd | by 3rd | 1st Choice | by 2nd | by 3rd |
| 10 | .81 | .89 | .93 | .69 | .83 | .89 |
| 20 | .82 | .89 | .94 | .66 | .78 | .85 |
| 30 | .89 | .94 | .97 | .68 | .84 | .89 |
| 40 | .86 | .93 | .98 | .67 | .79 | .86 |

## VII. CONCLUSION.

Some of the stochastic properties of the D-Wave useful for computation have been characterized, the qubit as a stochastic binary unit and the stochastic coupling of the qubits. It was shown how the architecture of the D-Wave fits very closely to the properties necessary to implement neural networks, the Hopfield network and Boltzmann machine in particular. It was also shown how the D-Wave can be used to implement a many layered neural network where the layers of the network utilizing the D-Wave can be of any size (not limited by the size of the D-Wave). A neural network was implemented using the D-Wave to recognize hand written digits demonstrating that such a network can be implemented using the D-Wave and performs well. Thus the goal of showing that it is feasible to implement neural networks for machine learning using the D-Wave has been satisfied.

REFERENCES

[1] G. E. Santoro and E. Tosatti, "Optimization using quantum mechanics: quantum annealing through adiabatic evolution", J. Phys. A 39, R393 (2006)".
[2] Edward Farhi, Jeffrey Goldstone, et al., "Quantum Computation by Adiabatic Evolution", MIT CTP #2936.
[3] J. J. Hopfield, Neural networks and physical systems with emergent collective computational abilities, *Proceedings of the National Academy of Sciences of the USA*, vol. 79 no. 8 pp. 2554–2558, April 1982.
[4] Geoffrey E. Hinton and Terrence J. Sejnowski, Analyzing Cooperative Computation, *Proceedings of the 5th Annual Congress of the Cognitive Science Society*, Rochester, New York, May 1983.
[5] Geoffrey E. Hinton and Terrence J. Sejnowski, Optimal Perceptual Inference, *Proceedings of the IEEE conference on Computer Vision and Pattern Recognition (CVPR)*, pages 448–453, IEEE Computer Society, Washington, D.C., June 1983.
[6] M. A. Carreira-Perpinan and G. E. Hinton. On contrastive divergence (CD) learning, *Proceedings of Tenth International Workshop on Artificial Intelligence and Statistics*, pg 33-40, Barbados, Jan. 2005.
[7] Denil, M. and de Freitas, N. Toward the implementation of a quantum RBM. In NIPS Deep Learning and Unsupervised Feature Learning Workshop, 2011.
[8] Dumoulin, V., Goodfellow, I. J., Courville, A., and Bengio, Y., On the Challenges of Physical Implementations of RBMs, Dec 2013, arXiv:1312.5258.